# A new superconducting open-framework allotrope of silicon at ambient pressure

Ha-Jun Sung[1], W. H. Han[1], In-Ho Lee[2], and K. J. Chang[1*]

[1]Department of Physics, Korea Advanced Institute of Science and Technology, Daejeon 34141, Korea

[2]Korea Research Institute of Standards and Science, Daejeon 34113, Korea

**Abstract**

Diamond Si is a semiconductor with an indirect band gap that is the basis of modern semiconductor technology. Although many metastable forms of Si were observed using diamond anvil cells for compression and chemical precursors for synthesis, no metallic phase at ambient conditions has been reported thus far. Here we report the prediction of pure metallic Si allotropes with open channels at ambient pressure, unlike a cubic diamond structure in covalent bonding networks. The metallic phase termed $P6/m$-$Si_6$ can be obtained by removing Na after pressure release from a novel Na-Si clathrate called $P6/m$-$NaSi_6$, which is discovered through first-principles study at high pressure. We confirm that both $P6/m$-$NaSi_6$ and $P6/m$-$Si_6$ are stable and superconducting with the critical temperatures of about 13 and 12 K at ambient pressure, respectively. The discovery of new Na-Si and Si clathrate structures presents the possibility of exploring new exotic allotropes useful for Si-based devices.

[*]Corresponding author address: kjchang@kaist.ac.kr

Silicon is an abundant, non-toxic element in the Earth's crust and widely used as the key material for electronic devices such as integrated circuits and solar cells. At ambient conditions, Si crystallizes in a cubic diamond structure and also exists in numerous metastable forms such as amorphous Si, porous Si, Si nanostructures (nanowires and silicene), and Si clathrates. It is known that Si forms various metallic phases under compression, following the transition sequence from cubic diamond (*cd*-Si, Si-I) to *β*-Sn (Si-II) to *Imma*-Si (Si-XI) to simple hexagonal (*sh*-Si, Si-V) to *Cmca*-Si (Si-VI) to hexagonal close-packed (Si-VII) to face-centered cubic (Si-X) [1−4]. Upon slow pressure release from the *β*-Sn phase, Si does not recover the cubic diamond phase but instead transforms to other metastable phases with distorted tetrahedral bonds, such as rhombohedral R8 (Si-XII) and body-centered BC8 (Si-III) phases [5,6]. Subsequent thermal annealing of BC8-Si leads to the formation of hexagonal diamond Si (Si-IV) [5,6]. Upon fast release of pressure, two metastable phases referred to as Si-VIII and Si-IX were obtained, but their crystal structures are not identified yet [7]. Besides that, an unidentified Si-XIII phase was discovered upon thermal annealing of the R8/BC8 mixture made by indentation [8]. Recently, based on the combined study of *ab initio* calculations and microexplosion experiments, two new structures called BT8 and ST12 were identified on Si surface exposed to ultrashort laser pulses [9].

Another route for synthesizing other Si allotropes is to use appropriate chemical precursors followed by physical and chemical manipulations. In this process, thermodynamically stable precursors can be produced under pressure and may recover their stability to ambient conditions. In Na-Si systems, the type-II Si clathrate ($Fd\bar{3}m$-Si$_{136}$) with cages has been synthesized by removing Na from the precursor of type-II Na$_{24}$Si$_{136}$ clathrate [10,11], while Na has not been successfully removed from the cages of type-I Na$_8$Si$_{46}$ clathrate. Recently, a new Na$_4$Si$_{24}$ clathrate was made at high pressure, and its crystal structure was maintained at ambient pressure [12]. Through a thermal degassing process, a new Si$_{24}$ clathrate with open channels, hereafter referred to as *Cmcm*-Si$_6$, was obtained from the Na$_4$Si$_{24}$ precursor, possessing a quasi-direct band gap near 1.3 eV [13]. Although the reported metastable Si allotropes (Si-IV, BC8, R8, ST12, BT8, $Fd\bar{3}m$-Si$_{136}$, and *Cmcm*-Si$_6$) can persist to ambient conditions, all of them are semiconducting [5,6,9,10,11,13], and metallic Si phases at ambient conditions have not been reported to date. Theoretically, BC8-Si was predicted to be a semimetal [14−16], but it has been recently confirmed as a narrow gap semiconductor from optical, electrical, and heat capacity measurements on phase-pure bulk samples [17]. The superconductivity of Si has been observed in heavily B-doped cubic diamond phase [18] and *β*-Sn and simple hexagonal phases under pressure [19−21]. However, the high pressure metallic phases do not maintain their crystal structures after the release of pressure [5−7].

Here we report the discovery of a new Na-Si clathrate structure, which can be synthesized at high pressure and used as a precursor to obtain a novel superconducting Si phase at ambient pressure. The new clathrate structure, termed *P*6*/m*-NaSi$_6$, contains open channels embedded in a simple hexagonal Si lattice and guest Na atoms filling the channels. The enthalpy of *P*6*/m*-NaSi$_6$ is lower than that of the recently discovered Na$_4$Si$_{24}$ clathrate above 12.4 GPa, and its stability is verified by phonon spectra and molecular dynamics simulations. We find that a novel metallic phase termed *P*6/*m*-Si$_6$ can be obtained through the thermal degassing from *P*6/*m*-NaSi$_6$, and both *P*6/*m*-NaSi$_6$ and *P*6/*m*-Si$_6$ exhibit the superconducting behavior at ambient pressure.

We explored new Na-Si clathrates at high pressure using an *ab initio* evolutionary crystal structure search method [22]. The enthalpy minimization and electronic structure calculations were performed within the framework of density functional theory (full details in Supplemental Material [23]). We considered Na-Si systems with various composition ratios, $Na_xSi_y$ ($x:y$ = 2 : 1, 1 : 1, 1 : 2, 1 : 3, 1 : 4, 1 : 5, 1 : 6), each of which contains up to three formula units per supercell. The formation enthalpies of $Na_xSi_y$ are plotted for the stable and metastable configurations at 10 GPa and 20 GPa (Fig. 1a). The configurations on the convex hull are enthalpically stable against the decomposition into Na metal and *cd*-Si. At 10 GPa, NaSi with $x:y$ = 1 : 1 crystallizes in a $Na_4Si_4$ Zintl phase lying on the convex hull. For $x:y$ = 1 : 6, our crystal structure search method reproduces the clathrate structure of $Na_4Si_{24}$ in the orthorhombic *Cmcm* space group (hereafter referred to as *Cmcm*-$NaSi_6$), which has been recently discovered via high-pressure synthesis [12]. For $Na_2Si$ and $NaSi_2$ systems, we find the stable structures in the *P*6/*mmm* and $Fd\bar{3}m$ space groups (Fig. 1b), respectively, both of which have been theoretically proposed for Li-Si, Ba-Si, and Ca-Si compounds [24−26].

As pressure increases to 20 GPa, the Zintl phase becomes unstable against decomposition, and $NaSi_3$ forms a hexagonal layered structure in the *P*6/*mmm* space group, which consists of alternating hexagonal and triangular Si rings, similar to the low pressure phase predicted for $LiSi_3$ [26]. For $NaSi_6$ systems, we find a very distinctive clathrate structure in the hexagonal *P*6/*m* space group (Fig. 2a). The new $NaSi_6$ clathrate, termed *P6/m*-$NaSi_6$, maintains its crystal structure and symmetry even after pressure is released. At ambient pressure, *P6/m*-$NaSi_6$ has the equilibrium lattice parameters, $a = b = 7.038$ Å and $c = 2.573$ Å, and two inequivalent Wyckoff positions, 1*a* (0, 0, 0.5) and 6*j* (0.145, 0.710, 0) for the Na and Si atoms, respectively. The *P6/m*-$NaSi_6$ clathrate consists of sevenfold Si atoms and contains open channels along the crystallographic *c*-axis. The open channels are formed from six-membered Si rings. The intercalated Na atoms along the open channels are sandwiched between hexagonal Si layers.

If the Si atoms are added to fill the open channels after removing the Na atoms, the crystal structure is exactly the same as a simple hexagonal lattice, which is known as the stable metallic phase of Si at pressures of 13.2−39.2 GPa [1,2]. Although various clathrate forms have been studied for Li-, Ca-, and Ba-Si systems at high pressure [24−26], to our knowledge, the $P6/m$ structure of NaSi$_6$ has not been reported yet.

For the composition ratio of $x : y = 1 : 6$, the stability of $P6/m$-NaSi$_6$ was compared with the known Na-Si clathrates under pressure by calculating the enthalpy difference per atom defined with respect to a mixture of Zintl and $cd$-Si,

$$\Delta H = [H(\text{Na}_x\text{Si}_y) - xH(\text{NaSi}) - (y-x)H(cd\text{-Si})]/(x+y), \qquad (1)$$

where $H(\text{Na}_x\text{Si}_y)$, $H(\text{NaSi})$, and $H(cd\text{-Si})$ are the enthalpies of Na$_x$Si$_y$ clathrate, Zintl, and $cd$-Si, respectively. For the type-I (Na$_8$Si$_{46}$) and type-II (Na$_{24}$Si$_{136}$) clathrates with different stoichiometries from NaSi$_6$, we chose the (NaSi$_{5.75}$ + 0.25 $cd$-Si) and (NaSi$_{5.667}$ + 0.333 $cd$-Si) systems, respectively. The relative enthalpies are plotted as a function of pressure in Fig. 2b. The $Cmcm$-NaSi$_6$ clathrate is enthalpically more favorable than the type-I clathrate at 7.7 GPa, in good agreement with the previous study [12]. As pressure increases above 12.4 GPa, the $P6/m$-NaSi$_6$ clathrate becomes the most stable phase.

The thermal stability of $P6/m$-NaSi$_6$ was examined by carrying out first-principles isothermal-isobaric molecular dynamics ($NpT$-MD) simulations at 15 GPa and 1000 K. For a 2×2×6 supercell containing 144 Si atoms and 24 Na atoms, $P6/m$-NaSi$_6$ was found to be stable up to 100 ps (see Supplemental Material [23], Fig. 1). Furthermore, we calculated the full phonon spectra and found no imaginary phonon modes, confirming the dynamical stability of $P6/m$-NaSi$_6$ (Fig. 3a). Based on the enthalpy vs. pressure curves and $NpT$-MD simulations, we suggest that the synthesis of *P6/m*-NaSi$_6$ is possible under high temperature conditions above 12.4 GPa. In addition, to guide the experimental identification of *P6/m*-NaSi$_6$, we simulated the X-ray diffraction pattern at 12.4 GPa (see Supplemental Material [23], Fig. 2).

When pressure is released from the *P6/m* phase of NaSi$_6$, one can remove the Na atoms from the open channels using a thermal degassing process and subsequently obtain a new Si allotrope in the space group *P6/m*, identical to the *sh* phase, except having open channels along the *c*-axis (Fig. 2a). Experiments have shown that the Na atoms can be removed by exposing Na-Si clathrates to elevated temperatures [10,11,13]. This degassing process occurs at temperatures 623−648 K for the type-II clathrate with polyhedral cavities [10]. In *Cmcm*-NaSi$_6$, the Na removal leading to the *Cmcm*-Si$_6$ allotrope was achieved at a lower temperature of 400 K due to the presence of open channels [13]. Since *P6/m*-NaSi$_6$ has open channels, the degassing process can also remove the Na atoms at relatively low temperatures. Using a climbing image nudged elastic band method [27], we investigated the migration of Na along the open channels for both the *Cmcm*- and *P6/m*-NaSi$_6$ allotropes and estimated the energy barriers for Na migration to be 0.65 and 0.48 eV, respectively (see Supplemental Material [23], Fig. 3). The lower energy barrier of *P6/m*-NaSi$_6$ is because the large open channels are connected in a straight line to weaken the interaction between the guest and its surrounding Si atoms. Our results suggest that the *P6/m*-Si$_6$ allotrope can be formed at temperatures around 400 K through the degassing process, similar to *Cmcm*-Si$_6$. The lattice parameters of *P6/m*-Si$_6$ after the Na removal are reduced to $a = b = 6.854$ Å and $c = 2.478$ Å. Since the *P6/m*-Si$_6$ structure is composed of highly coordinated Si atoms, this high-density phase is metallic, unlike the low-density *Cmcm*-Si$_6$ allotrope having a semiconducting band gap due to fourfold coordinated Si atoms.

The electronic band structures of $P6/m$-NaSi$_6$ and $P6/m$-Si$_6$ are compared in Fig. 3b. In the type-I, type-II, and $Cmcm$-NaSi$_6$ clathrates, all Si atoms are tetrahedrally bonded within an $sp^3$ covalent network, as in the diamond phase. Thus, these clathrates exhibit the semiconducting band structure after the guest atoms are removed [10,11,13]. In $P6/m$-NaSi$_6$, although the Na degassing process lowers the Fermi level, the metallic band structure persists for $P6/m$-Si$_6$. The $sh$ phase of Si was observed to be superconducting with a critical temperature of 8.2 K at 15.2 GPa [20,21]. Similar to $sh$-Si, both $P6/m$-NaSi$_6$ and $P6/m$-Si$_6$ have the metallic bonding network and high electronic density of states as ingredients for superconductivity.

To explore the superconductivity of $P6/m$-NaSi$_6$ and $P6/m$-Si$_6$, the electron-phonon (EP) coupling constant ($\lambda$) was calculated from density-functional perturbation theory (full details in Supplemental Material [23]). The Allen−Dynes modified McMillan equation was used to estimate the superconducting critical temperature ($T_c$) [28,29], with choosing the Coulomb pseudopotential of $\mu^* = 0.1$. The logarithmic average of phonon frequencies ($\omega_{\log}$) and frequency-dependent EP coupling $\lambda(\omega)$ were obtained from the Eliashberg spectral function $\alpha^2 F(\omega)$ (Fig. 3a). The results for $\lambda$, $\omega_{\log}$, and $T_c$ are shown for $sh$-Si, $P6/m$-NaSi$_6$, and $P6/m$-Si$_6$ in Table I. The superconducting temperature of $sh$-Si is estimated to be 7.9 K at 15 GPa, in good agreement with the reported value [20,21]. We find that both the $P6/m$-NaSi$_6$ and $P6/m$-Si$_6$ allotropes are superconducting with the critical temperatures of 13.1 and 12.2 K at zero pressure, respectively. These $T_c$ values are higher than those (4−8 K) observed for the high-pressure $\beta$-Sn and $sh$ phases of Si [19−21] and the type-I clathrate of (Na,Ba)$_x$Si$_{46}$ and Ba$_8$Si$_{46}$ [30,31]. As pressure increases to 15 GPa, the critical temperature decreases to 4.0

K for $P6/m$-NaSi$_6$ and 5.3 K for $P6/m$-Si$_6$ (Fig. 4a). The pressure coefficients of $T_c$ are −0.61 and −0.46 K/GPa for $P6/m$-NaSi$_6$ and $P6/m$-Si$_6$, respectively, similar to the measured value of −0.52 K/GPa for $sh$-Si [20].

In both $P6/m$-NaSi$_6$ and $P6/m$-Si$_6$, there are four notable Raman active modes at the Γ point (Fig. 3a), $A_g^1$ and $A_g^2$ modes associated with torsional and breathing motions of Si atoms around the open channels, respectively, and $E_{2g}^1$ and $E_{2g}^2$ modes related to stretching and bending distortions of in-plane Si-Si bonds (see Supplemental Material [23], Fig. 4). In $P6/m$-NaSi$_6$, the interaction of the $E_{2g}^1$ optical modes with electrons is strong along Γ−A, A−L, and A−H in the Brillouin zone (Fig. 3a). However, there are no specific phonon modes with dominant contributions to the EP coupling. The density of states at the Fermi level, $N(0)$, is higher in $P6/m$-NaSi$_6$ than $P6/m$-Si$_6$ (Table I), and therefore the higher $T_c$ of $P6/m$-NaSi$_6$ than $P6/m$-Si$_6$ at zero pressure is mainly related to $N(0)$. Since the EP coupling is weakened and $N(0)$ decreases under pressure (see Supplemental Material [23], Fig. 5), both $P6/m$-NaSi$_6$ and $P6/m$-Si$_6$ show the decreasing pressure behavior of $T_c$. In $P6/m$-Si$_6$, the transverse acoustic (TA) modes around the A point are softened under pressure, unlike $P6/m$-NaSi$_6$ (Fig. 4b). Since the crystal volume is reduced after the Na removal, the phonon modes are more rigid in $P6/m$-Si$_6$ than $P6/m$-NaSi$_6$ (Fig. 3a). Although $\omega_{\log}$ is higher in $P6/m$-Si$_6$ at zero pressure, its value becomes lower than that of $P6/m$-NaSi$_6$ as pressure increases above 7 GPa (see Supplemental Material [23], Fig. 5). Due to the soft TA modes, the EP coupling of $P6/m$-Si$_6$ appears strongly along Γ−A, A−L, and A−H and is strengthened as pressure increases (Fig. 3a), resulting in a smaller pressure coefficient of $T_c$.

Finally, we discuss the stability of $P6/m$-Si$_6$ at ambient pressure. The total energy of $P6/m$-Si$_6$ is calculated to be about 0.35 eV/atom higher than that of $cd$-Si (Fig. 4c). Since $P6/m$-Si$_6$ is less stable by 0.26 eV/atom than *Cmcm*-Si$_6$, its thermal stability is more sensitive to temperature. Nonetheless, we confirmed the thermal and dynamical stability of *P*6/*m*-Si$_6$ up to 400 K at ambient pressure from the phonon spectra and *NpT*-MD simulations (see Supplemental Material [23], Figs. 1 and 6). When heated above 400 K, *P*6/*m*-Si$_6$ forms an amorphous-like disordered structure and then may decompose or turn to the most stable *cd*-Si phase. In the type-II Na$_{24}$Si$_{136}$ clathrate, it is known that the guest atoms frustrate a transition into another phase at high pressure [32]. Similarly, in *P*6/*m*-NaSi$_6$, the guest atoms residing along the open channels suppress the motions of the Si layers and thus maintain the structural stability even at high pressure. In *P*6/*m*-Si$_6$, the TA modes at the A point are associated with the displacements of the hexagonal Si layers along the [110] and [1$\bar{1}$0] directions (Fig. 4b). Since two adjacent layers move in opposite directions, the open channels can be broken by large atomic displacements. Although the instability of the TA modes leads to a transition to the hexagonal close-packed (hcp) structure under pressure, the energy versus volume curves show that *P*6/*m*-Si$_6$ is likely to transform to a metallic *Cmca*-Si phase (at 6.6 GPa) prior to the transition to hcp (at 12.1 GPa) (Fig. 4c).

In conclusion, we have discovered a new *P*6/*m*-NaSi$_6$ clathrate which can be synthesized at high pressures above 12.4 GPa. The *P*6/*m*-NaSi$_6$ structure is characterized by a simple hexagonal Si lattice containing open channels that are filled with the Na atoms. The stability of *P*6/*m*-NaSi$_6$ down to zero pressure was verified by phonon spectra and molecular dynamics simulations. We propose that, using the *P*6/*m*-NaSi$_6$ clathrate as a precursor, the Na degassing process can produce a pure metallic Si phase at ambient pressure, termed *P*6/*m*-Si$_6$. Both the *P*6/*m*-NaSi$_6$ and *P*6/*m*-Si$_6$ allotropes are predicted to be superconducting with the critical temperatures of 13.1 and 12.2 K at zero pressure, respectively. The discovery of the novel superconducting Si phase not only provides a new paradigm for understanding the energy landscape of Si but offers an opportunity to explore further exotic Si allotropes to be realized for new Si-based devices.

**Acknowledgments** This work was supported by Samsung Science and Technology Foundation under Grant No. SSTF-BA1401-08.

TABLE 1. Crystal volumes ($V$ in Å$^3$ per Si-atom), densities of states at the Fermi level [$N(0)$ in states/Ry/Si-atom/spin], EP coupling constants ($\lambda$), logarithmic averages of phonon frequencies ($\omega_{\log}$ in K), and superconducting critical temperatures ($T_c$ in K) for $P6/m$-NaSi$_6$, $P6/m$-Si$_6$, and simple hexagonal Si ($sh$-Si).

| Allotrope | Pressure (GPa) | $V$ | $N(0)$ | $\lambda$ | $\omega_{\log}$ | $T_c$ |
|---|---|---|---|---|---|---|
| $P6/m$-NaSi$_6$ | 0 | 18.4 | 2.82 | 0.897 | 225 | 13.1 |
|  | 15 | 16.1 | 2.60 | 0.498 | 331 | 4.0 |
| $P6/m$-Si$_6$ | 0 | 16.8 | 2.40 | 0.799 | 263 | 12.2 |
|  | 15 | 14.8 | 2.22 | 0.556 | 299 | 5.3 |
| $sh$-Si | 15 | 13.5 | 2.44 | 0.660 | 250 | 7.9 |

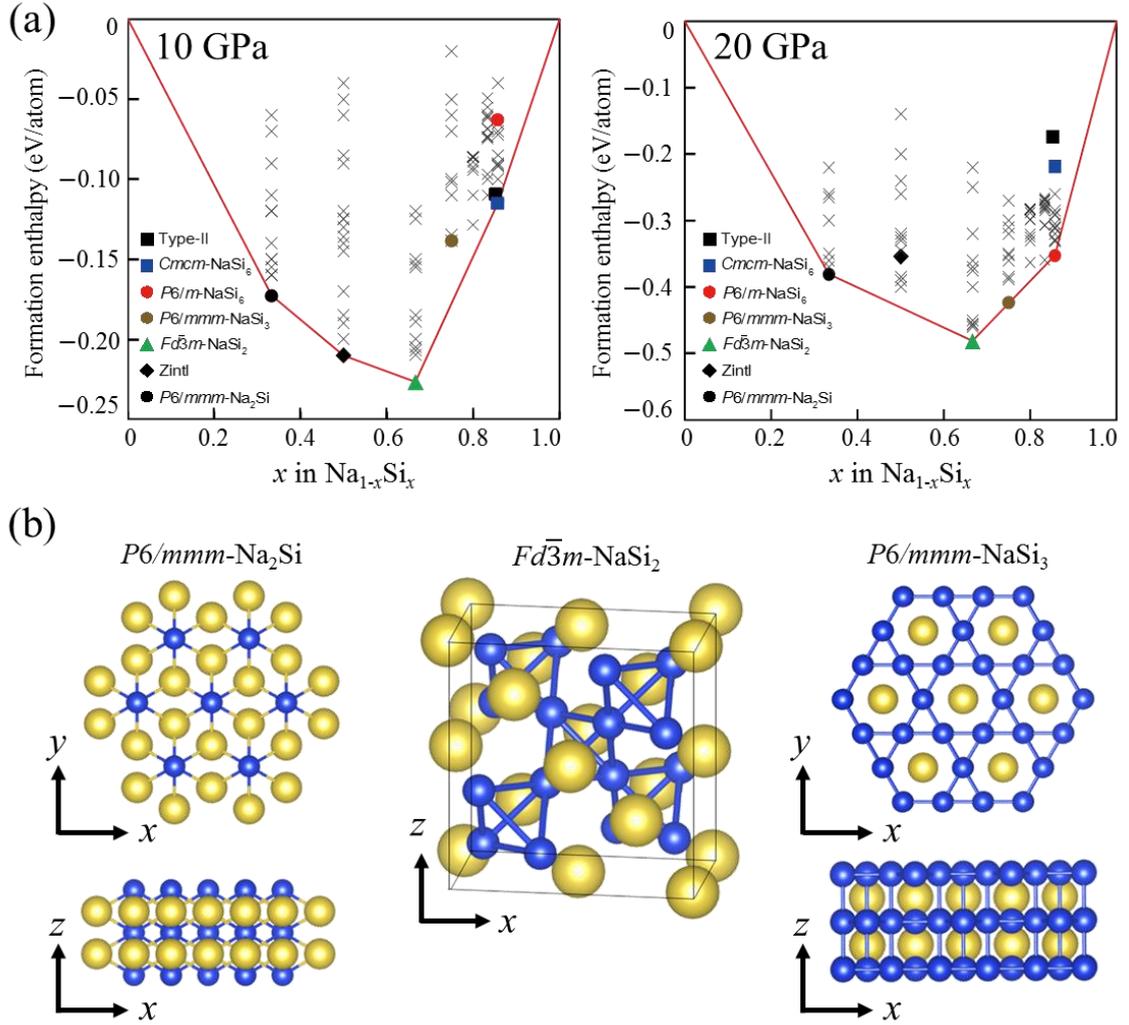

FIG. 1. (a) Formation enthalpies of stable and metastable configurations for Na$_{1-x}$Si$_x$ with various composition ratios at pressures of 10 GPa (left) and 20 GPa (right). The configurations on the convex hull are marked by solid symbols and connected by red lines. (b) Atomic structures of $P6/mmm$-Na$_2$Si, $Fd\bar{3}m$-NaSi$_2$, and $P6/mmm$-NaSi$_3$ on the convex hull. The blue and yellow spheres represent Si and Na atoms, respectively.

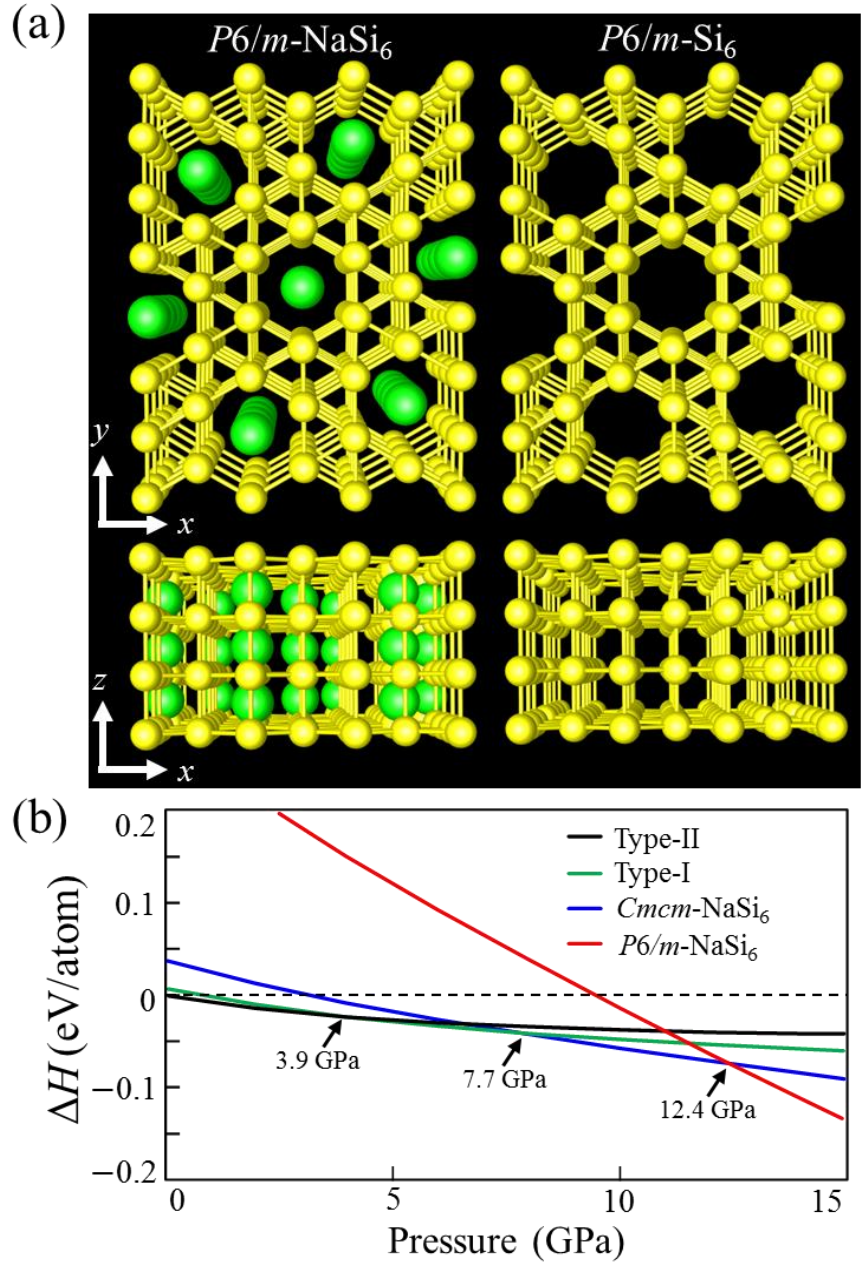

FIG. 2. (a) Top and side views of the atomic structures of $P6/m$-NaSi$_6$ (left) and $P6/m$-Si$_6$ (right) clathrates. In $P6/m$-NaSi$_6$, the open channels filled with Na atoms are embedded in the metallic bonding networks of Si atoms that are exactly the same as that of a simple hexagonal lattice except for the open channels. The blue and yellow spheres represent Si and Na atoms, respectively. (b) Relative enthalpies ($\Delta H$) of various clathrate structures with respect to a mixture of Zintl and $cd$-Si as a function of pressure.

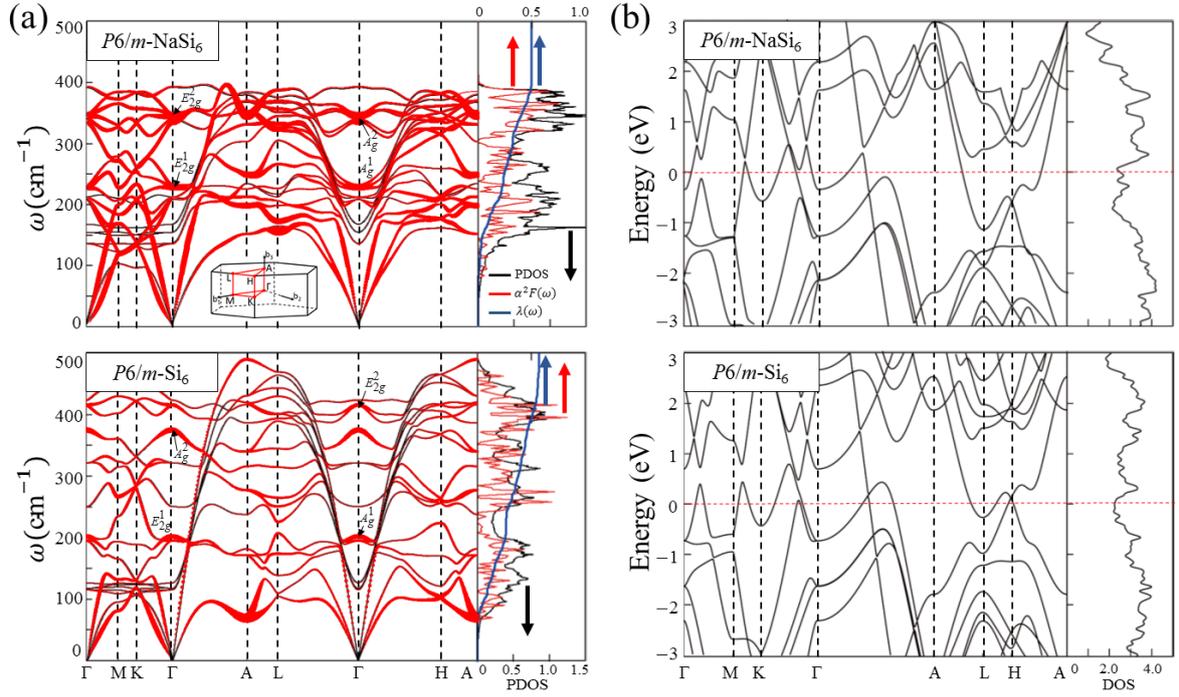

FIG. 3. (a) Phonon spectra, phonon densities of states (PDOS in arb. unit), Eliashberg spectral functions $\alpha^2F(\omega)$, and electron-phonon couplings $\lambda(\omega)$ for $P6/m$-NaSi$_6$ (top) and $P6/m$-Si$_6$ (bottom) at 15 GPa. The magnitude of $\lambda(\omega)$ is indicated by the thickness of the red curves. (b) Band structures and electronic densities of states (DOS in arb. unit) of $P6/m$-NaSi$_6$ (top) and $P6/m$-Si$_6$ (bottom) at 15 GPa, with the Fermi level set to zero.

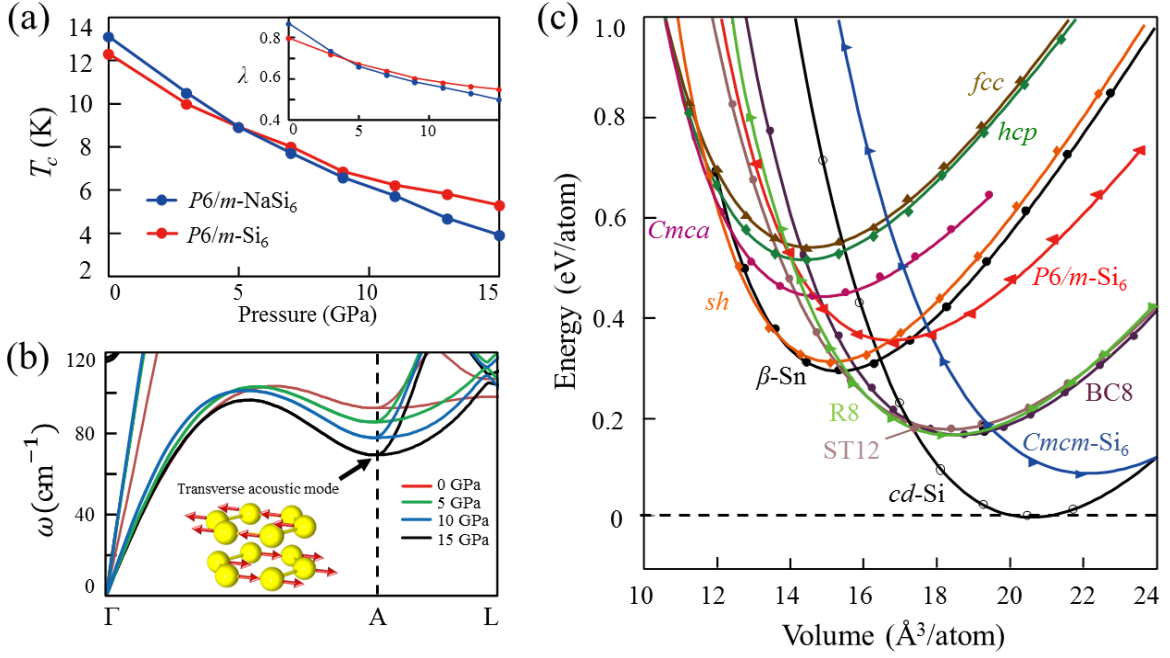

FIG. 4. (a) Pressure dependences of $T_c$ and $\lambda$ for $P6/m$-NaSi$_6$ and $P6/m$-Si$_6$. (b) Pressure variation of the low frequency modes of $P6/m$-Si$_6$ along the Γ-A-L line in the BZ. The balls represent the Si atoms around the open channels and the arrows stand for the displacements of the TA mode at the A point. (c) Energy vs. volume curves of various Si phases.

# Supplemental Material

# A new superconducting open-framework allotrope of silicon at ambient pressure

Ha-Jun Sung,[1] W. H. Han,[1] In-Ho Lee,[2] and K. J. Chang[1*]

[1]*Department of Physics, Korea Advanced Institute of Science and Technology, Daejeon 34141, Korea*

[2]*Korea Research Institute of Standards and Science, Daejeon 34113, Korea*

[*]Corresponding Author address: kjchang@kaist.ac.kr

## 1. Calculation method

For Na-Si systems with various composition ratios, low-enthalpy allotropes were explored using an *ab initio* evolutionary crystal structure search method, as implemented in the AMADEUS code [1]. In this method, the conformational space annealing (CSA) algorithm [2] for global optimization is combined with first-principles electronic structure calculations. Since distinct configurations are generated, the CSA algorithm is very efficient to search for both stable and metastable configurations. The number of configurations was set to 30 in the population size of the CSA, and the enthalpy was used to express the objective function that describes specific target properties for a given pressure. For each configuration, the enthalpy was minimized by performing density functional theory calculations which employed the functional form of Perdew, Burke, and Ernzerhof (PBE) for the exchange-correlation potential [3] and the projector augmented wave potentials [4], as implemented in the VASP code [5].

The wave functions were expanded in plane waves up to an energy cutoff of 600 eV. For Brillouin zone (BZ) integration, a *k*-point set was generated by a grid spacing of 0.15 Å$^{-1}$. The ionic coordinates and lattice parameters were fully optimized until the residual forces and stress tensors were less than 0.02 eV/Å and 1.5 kbar, respectively.

The phonon spectrum and EP coupling constant ($\lambda$) were obtained from first-principles density-functional perturbation theory as implemented in the QUANTUM ESPRESSO package [6]. The PBE exchange-correlation functional, ultrasoft pseudopotentials [7], and a plane wave basis set with an energy cutoff of 80 Ry were employed. The dynamical and electron-phonon interaction matrices were calculated using the 3×3×9 and 9×9×9 phonon-momentum grids for *P*6/*m*-Si$_6$ (*P*6/*m*-NaSi$_6$) and *sh*-Si, respectively. Then, denser 18×18×54 and 54×54×54 grids were adopted to perform the BZ integration in phonon-momentum space. The electron-phonon coupling $\lambda_{\mathbf{q}\nu}$ at a phonon wave vector **q** for a phonon mode $\nu$ is related to the phonon linewidth $\gamma_{\mathbf{q}\nu}$, $\lambda_{\mathbf{q}\nu}=\gamma_{\mathbf{q}\nu}/2\pi\omega_{\mathbf{q}\nu}^2 N(0)$, where $\omega_{\mathbf{q}\nu}$ is the phonon frequency. The phonon modes contributed to the electron-phonon coupling can be analyzed from the Eliashberg function defined as, $\alpha^2 F(\omega)=\frac{1}{2}\sum_\nu \int_{BZ} \frac{d\mathbf{q}}{\Omega_{BZ}} \lambda_{\mathbf{q}\nu}\omega_{\mathbf{q}\nu}\delta(\omega-\omega_{\mathbf{q}\nu})$, where $\Omega_{BZ}$ is the BZ volume. Then, the frequency-dependent electron-phonon coupling is obtained from $\lambda(\omega) = 2\int_0^\omega [\alpha^2 F(\omega')/\omega']d\omega'$ and the electron-phonon coupling constant is given by $\lambda = \lambda(\infty)$.

## 2. X-ray diffraction pattern simulations

To guide the experimental identification of *P*6/*m*-NaSi$_6$, X-ray diffraction (XRD) patterns were simulated at 12.4 GPa [8]. The simulated XRD patterns of *P*6/*m*-NaSi$_6$ show a very strong peak around 40.49°, corresponding to the (121) diffraction, as shown in Fig. 2(a). The other peaks (001), (101), (111), (201), and (211) correspond to 2θ = 36.03°, 39.24°, 45.01°, 47.72°, and 55.24°, respectively. The lattice parameters of $P6/m$-NaSi$_6$ and its unit cell volume decrease after the Na removal. Therefore, the diffraction peaks of $P6/m$-Si$_6$ slightly shift to the higher 2θ values [Fig. 2(b)]. In $P6/m$-Si$_6$, additional peaks (100), (110), and (200) appear at 2θ = 15.54°, 27.08°, and 31.37°, respectively.

## 3. Energy barriers for Na migration and stability confirmation

Figure 3(a) shows the migration pathway of a Na atom from site A to site B along the open channels in $P6/m$-NaSi$_6$ and $Cmcm$-NaSi$_6$. The migration barriers were calculated using the climbing image nudged elastic band (CI-NEB) method [9]. A 2×2×3 supercell containing one Na and 72 host Si atoms was adopted. For $Cmcm$-NaSi$_6$ and $P6/m$-NaSi$_6$, the saddle points were found to be located within the Si layer between the A and B sites, and the energy barriers were calculated to be 0.65 and 0.48 eV per Na atom, respectively, using the PBE functional for the exchange-correlation potential [10] [Fig. 3(b)]. The lower energy barrier for Na migration in $P6/m$-NaSi$_6$ is because the large open channels are connected in a straight line to weaken the interaction between the guest and its surrounding Si atoms. Based on the results, it is expected that the Na degassing process can occur at lower temperatures in $P6/m$-NaSi$_6$ than $Cmcm$-NaSi$_6$. The dynamical stability of $P6/m$-NaSi$_6$ and $P6/m$-Si$_6$ was confirmed by calculating the full phonon spectra at zero pressure (Fig. 6). Additionally, the thermal stability of $P6/m$-Si$_6$ was examined by carrying out isothermal-isobaric molecular dynamics ($NpT$-MD) simulations up to 100 ps at zero pressure and a temperature of 400 K [Fig. 1(b)].

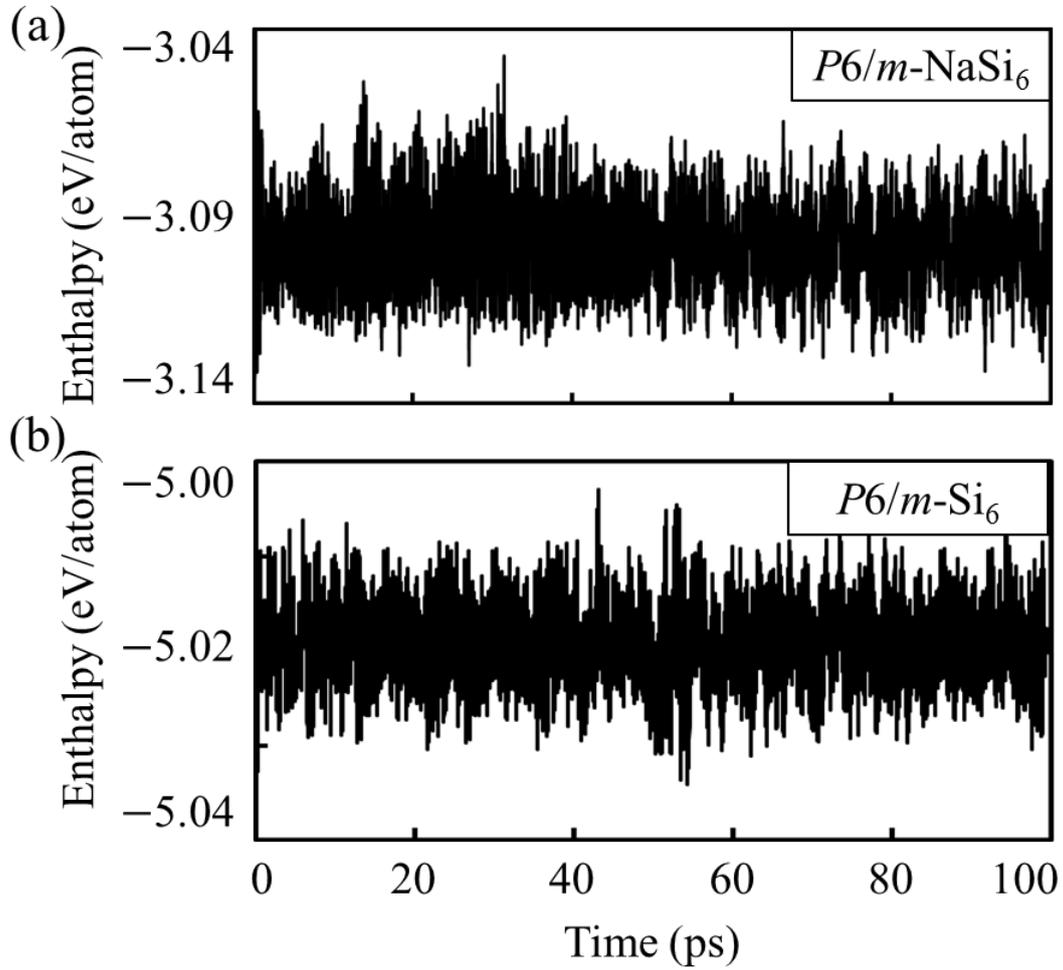

FIG. 1. Enthalpy fluctuations during *NpT*-MD simulations up to 100 ps for (a) *P*6/*m*-NaSi$_6$ at 15 GPa and 1000 K and (b) for *P*6/*m*-Si$_6$ at zero pressure and 400 K. For both *P*6/*m*-NaSi$_6$ and *P*6/*m*-Si$_6$, a $2\times2\times6$ supercell containing 144 host Si atoms is used.

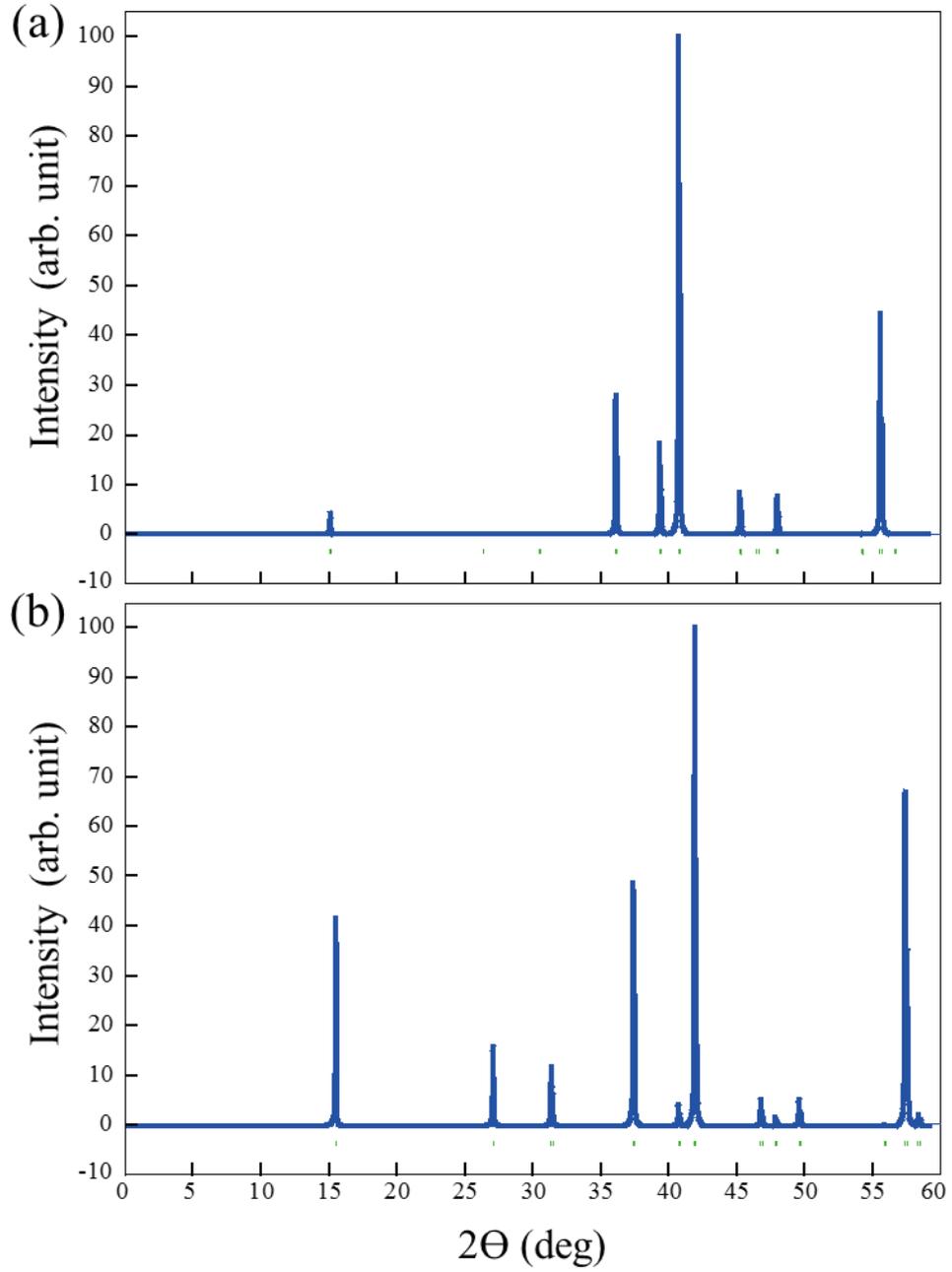

FIG. 2. The simulated X-ray diffraction patterns for (a) $P6/m$-NaSi$_6$ and (b) $P6/m$-Si$_6$ at 12.4 GPa. The lattice parameters of $P6/m$-NaSi$_6$ are $a = b = 6.762$ Å and $c = 2.485$ Å, whereas those of $P6/m$-Si$_6$ are $a = b = 6.578$ Å and $c = 2.403$ Å. A Cu source with the X-ray wavelength of 1.5406 Å is assumed.

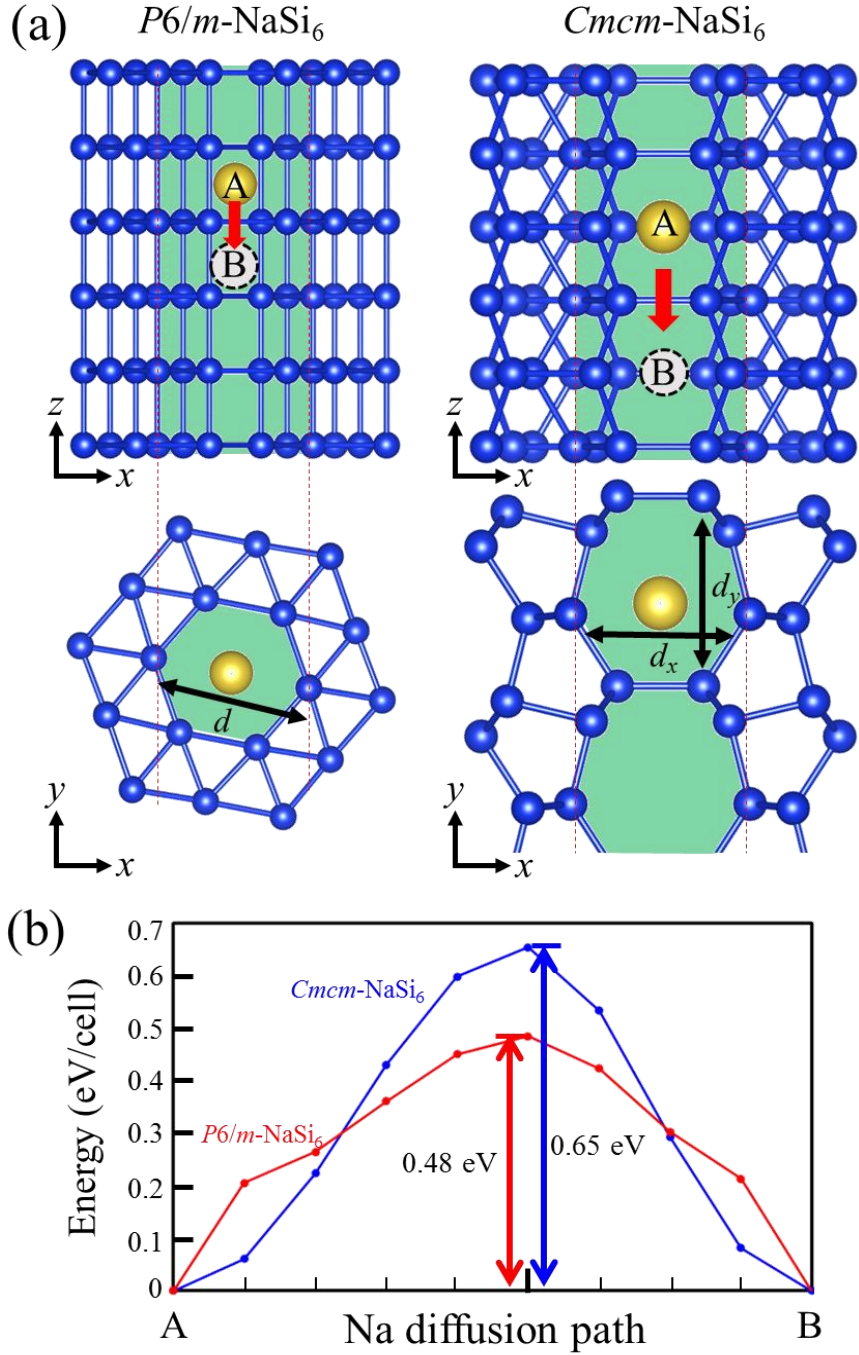

FIG. 3. (a) The migration pathways of Na atoms along the open channels in $P6/m$-NaSi$_6$ and $Cmcm$-NaSi$_6$. The size of open channels is $d = 5.27$ Å in $P6/m$-NaSi$_6$, and the channel sizes along the $x$ and $y$ directions are $d_x = 5.01$ Å and $d_y = 5.29$ Å in $Cmcm$-NaSi$_6$. (b) The energy barriers for Na migration along the open channels at zero pressure.

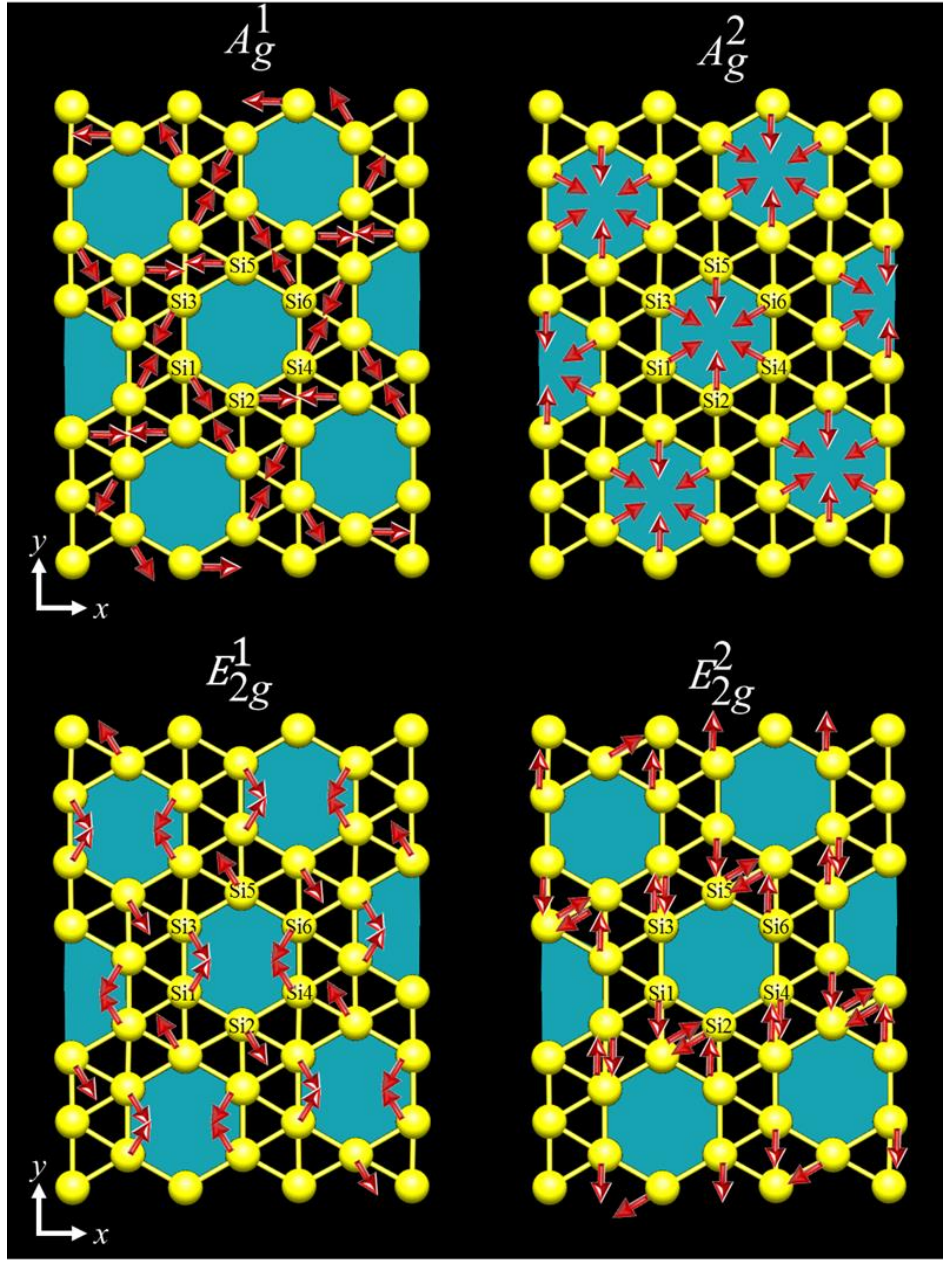

FIG. 4. Schematic illustrations of the atomic displacements of the Raman active $A_g^1$, $A_g^2$, $E_{2g}^1$, and $E_{2g}^2$ modes at the Γ point in *P*6/*m*-NaSi$_6$ and *P*6/*m*-Si$_6$. Blue hexagons represent the open channels in top view. The $A_g^1$ and $A_g^2$ modes are associated with torsional and breathing motions of Si atoms around the open channels, respectively. The $E^1_{2g}$ and $E^2_{2g}$ modes are related to stretching and bending distortions of in-plane Si-Si bonds.

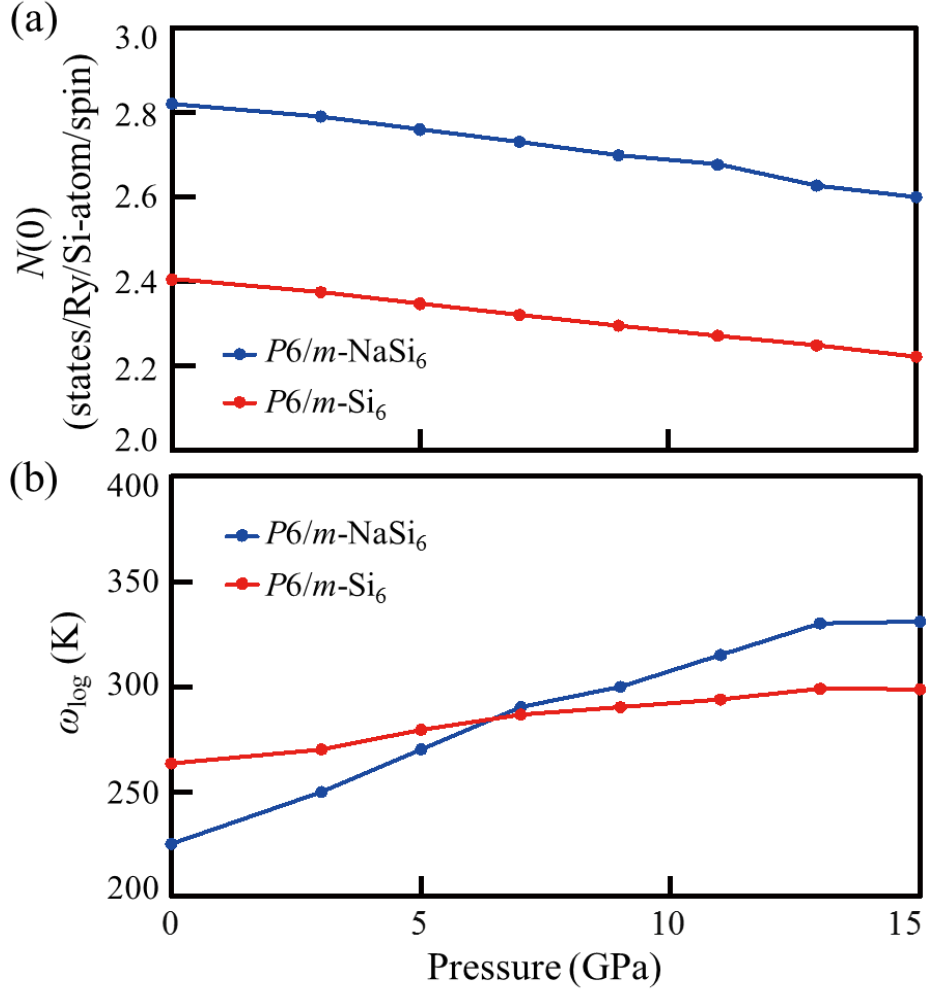

FIG. 5. (a) The electronic density of states at the Fermi level, $N(0)$, and (b) the logarithmic average of the phonon frequency, $\omega_{\log}$, are plotted as a function of pressure for $P6/m$-NaSi$_6$ and $P6/m$-Si$_6$.

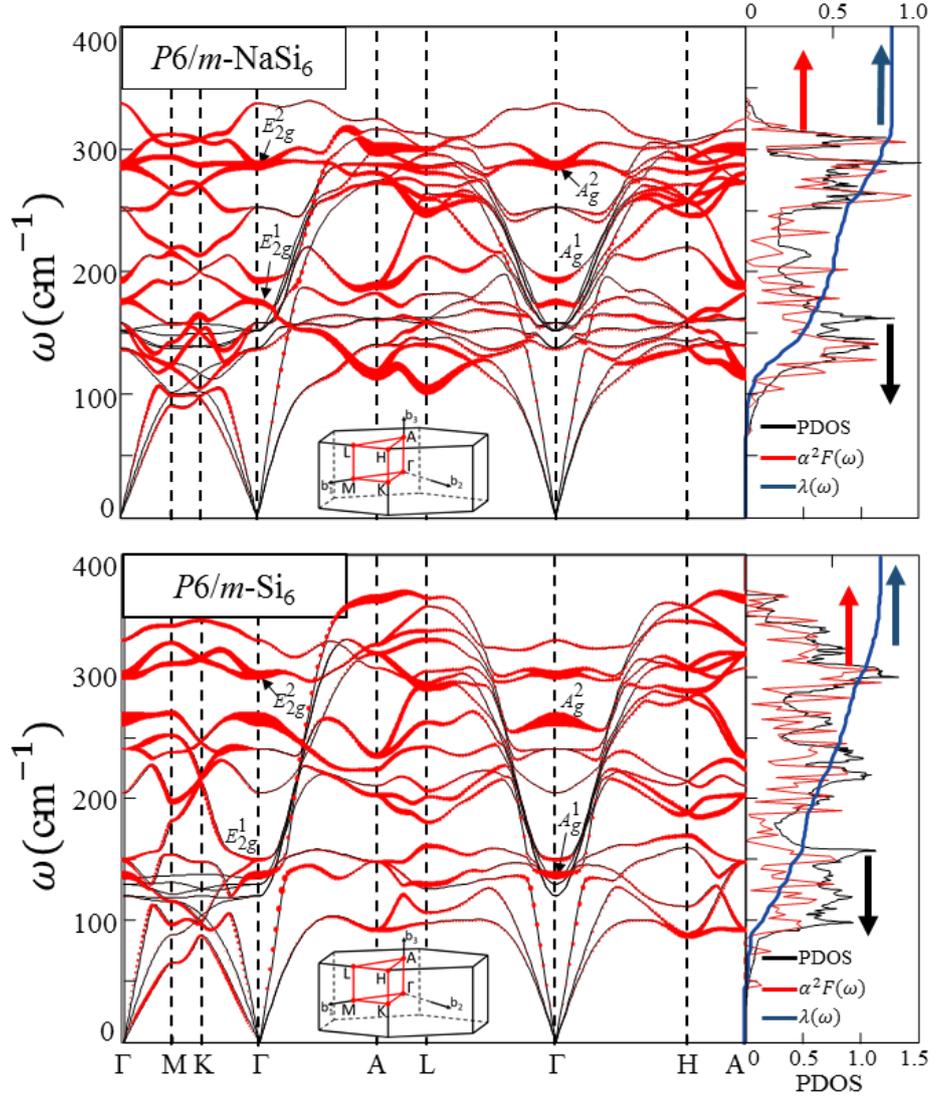

FIG. 6. The phonon spectra, phonon densities of states (PDOS in arb. unit), Eliashberg spectral functions $\alpha^2 F(\omega)$, and electron-phonon couplings $\lambda(\omega)$ for $P6/m$-NaSi$_6$ (top) and $P6/m$-Si$_6$ (bottom) at zero pressure. The magnitude of $\lambda(\omega)$ is indicated by the thickness of the red curves.